\documentclass[aps,prl,superscriptaddress,floatfix,showpacs,twocolumn]{revtex4}
\usepackage{verbatim}

\usepackage{amsmath}
\usepackage{amsfonts}
\usepackage{amssymb}
\usepackage{graphicx}
\usepackage{bm}
\usepackage{color}
\usepackage{epsf}
\usepackage[hypertex]{hyperref}
\usepackage{psfrag}
\def\bea{\begin{eqnarray}}
\def\eea{\end{eqnarray}}
\def\beas{\begin{eqnarray*}}
\def\eeas{\end{eqnarray*}}
\def\beqas{\begin{eqnarray*}}
\def\eqas{\end{eqnarray*}}
\def\beq{\begin{equation}}
\def\eeq{\end{equation}}
\def\beqd{\begin{displaymath}}
\def\eeqd{\end{displaymath}}
\def\eqd{\end{displaymath}}

\def\slashchar#1{\setbox0=\hbox{$#1$}
   \dimen0=\wd0
   \setbox1=\hbox{/} \dimen1=\wd1
   \ifdim\dimen0>\dimen1
      \rlap{\hbox to \dimen0{\hfil/\hfil}}
      #1
   \else\begin{eqnarray}
      \rlap{\hbox to \dimen1{\hfil$#1$\hfil}}
      /
   \fi}

\begin{document}
\title
{On timelike and spacelike hard exclusive reactions. }
\author{D.~M\"uller}
\affiliation{
Institut f\"ur Theoretische Physik II, Ruhr-Universit\"at Bochum, D-44780 Bochum, Germany}
\author{ B.~Pire}
\affiliation{ CPhT, \'Ecole Polytechnique,
CNRS, F-91128 Palaiseau,     France }
\author{ L.~Szymanowski}
\author{J.~Wagner}
\affiliation{National Center for Nuclear Research (NCBJ), Warsaw, Poland}

%\date{\today}

\begin{abstract}

\noindent
We show to next-to-leading order accuracy in the strong coupling $\alpha_s$ how the collinear factorization properties of QCD in the generalized Bjorken regime relate exclusive amplitudes for spacelike and timelike hadronic processes. This yields simple  space--to--timelike  relations linking
%on the one hand
the amplitudes for electroproduction of a photon or meson to those for photo- or meso-production of a lepton pair.
%on the other hand.
These relations constitute a new  test of the relevance of  leading twist analyzes of experimental data.
%in terms of generalized parton distributions.
\end{abstract}
\pacs{13.88.+e,13.85.Qk,12.38.Bx}

\maketitle

In the traditional collinear factorization framework the scattering amplitude for  exclusive processes
has been shown \cite{DA, historyofDVCS,Collins:1996fb} to factorize in specific kinematical regions, provided a large scale controls the separation of short distance dominated partonic subprocesses and long distance hadronic matrix elements. This large scale may come from a spacelike momentum exchange, as in hard leptoproduction processes, or from a timelike momentum as in electron-positron annihilation or lepton pair production.

The complementarity of spacelike and timelike processes has been much used in inclusive reactions to understand in detail parton distribution  and parton fragmentation functions, in particular  through deep inelastic leptoproduction and Drell--Yan processes in hadron reactions. In the realm of exclusive reaction, much work has been devoted to the electromagnetic form factors. In particular, the spacelike and timelike meson form factors were analyzed in great details  in Ref. \cite{BRS} .

Analyticity of the factorized amplitude is the basic
property that allows us to derive the new relations Eqs. \ref{fullrelation}, \ref{relationC-SL2TL_2}   at the
heart of our paper. Analyticity, which is a consequence of causality in relativistic field theory, and factorization
of short distance vs long distance properties, are common tools in many fields of theoretical physics.
Our instance is to our knowledge the first case where they are put together to obtain useful relations between observables.

\begin{figure}
%\begin{picture}(78,58)
%\put(0,28){\includegraphics[width=3.4cm]{figDVCS.eps}}
%\put(48,30){\includegraphics[width=3.4cm]{figTCS.eps}
%\put(0,0){\includegraphics[width=3.4cm]{figDVCSpi.eps}}
%\put(48,0){\includegraphics[width=3.4cm]{figPiTCS.eps}}
%\end{picture}
\includegraphics[width=3.6cm]{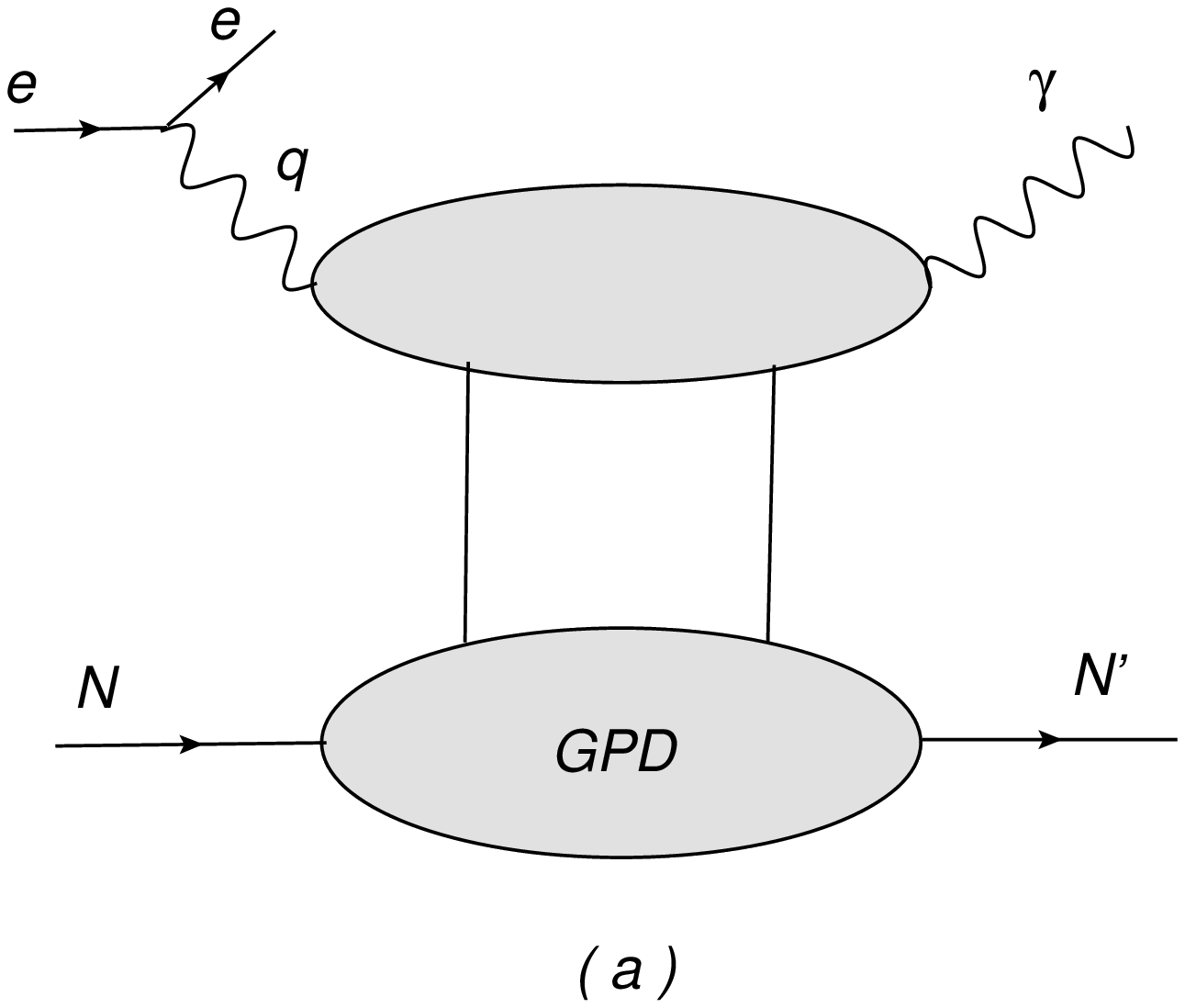}~~~\includegraphics[width=3.9cm]{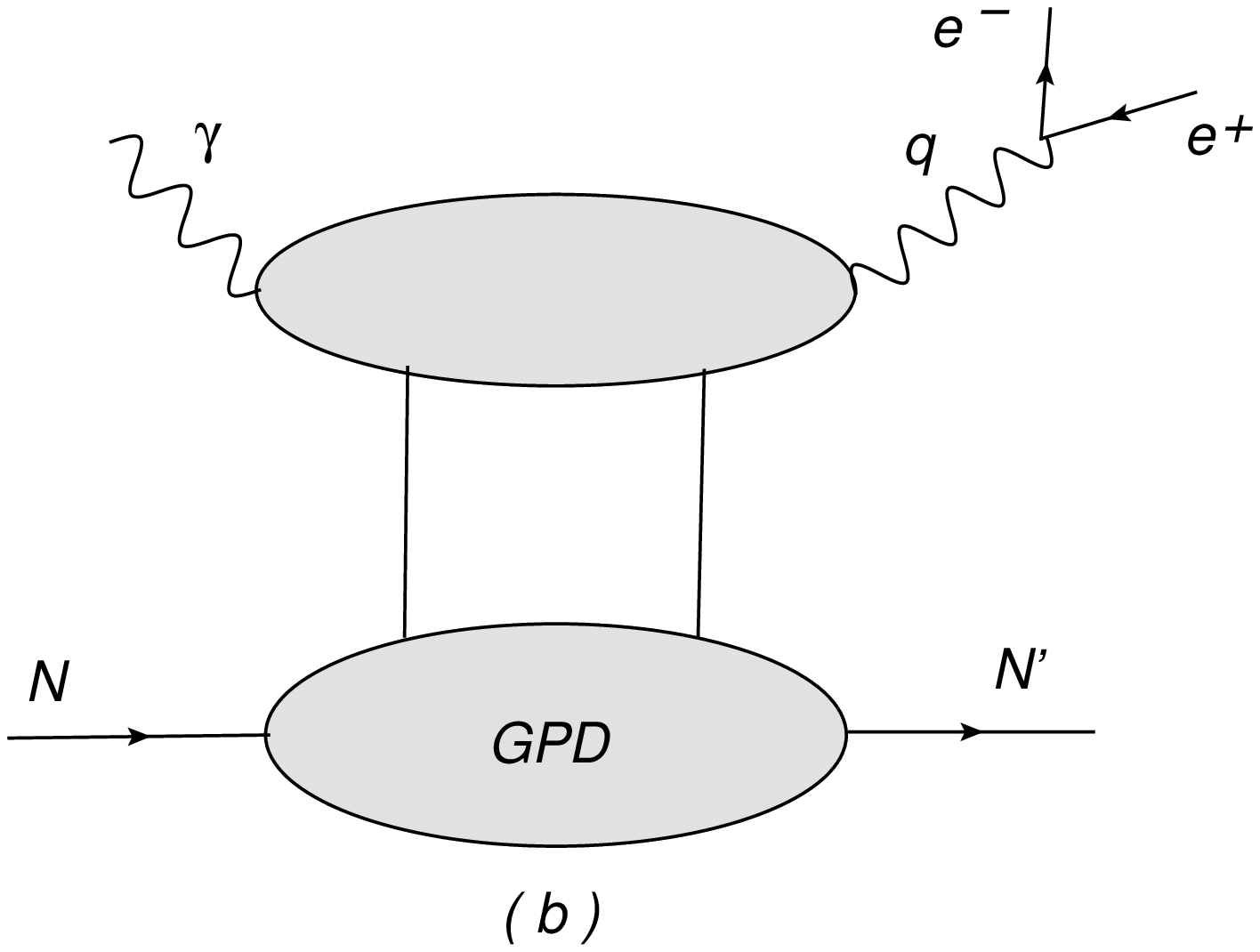}\\
\includegraphics[width=3.8cm]{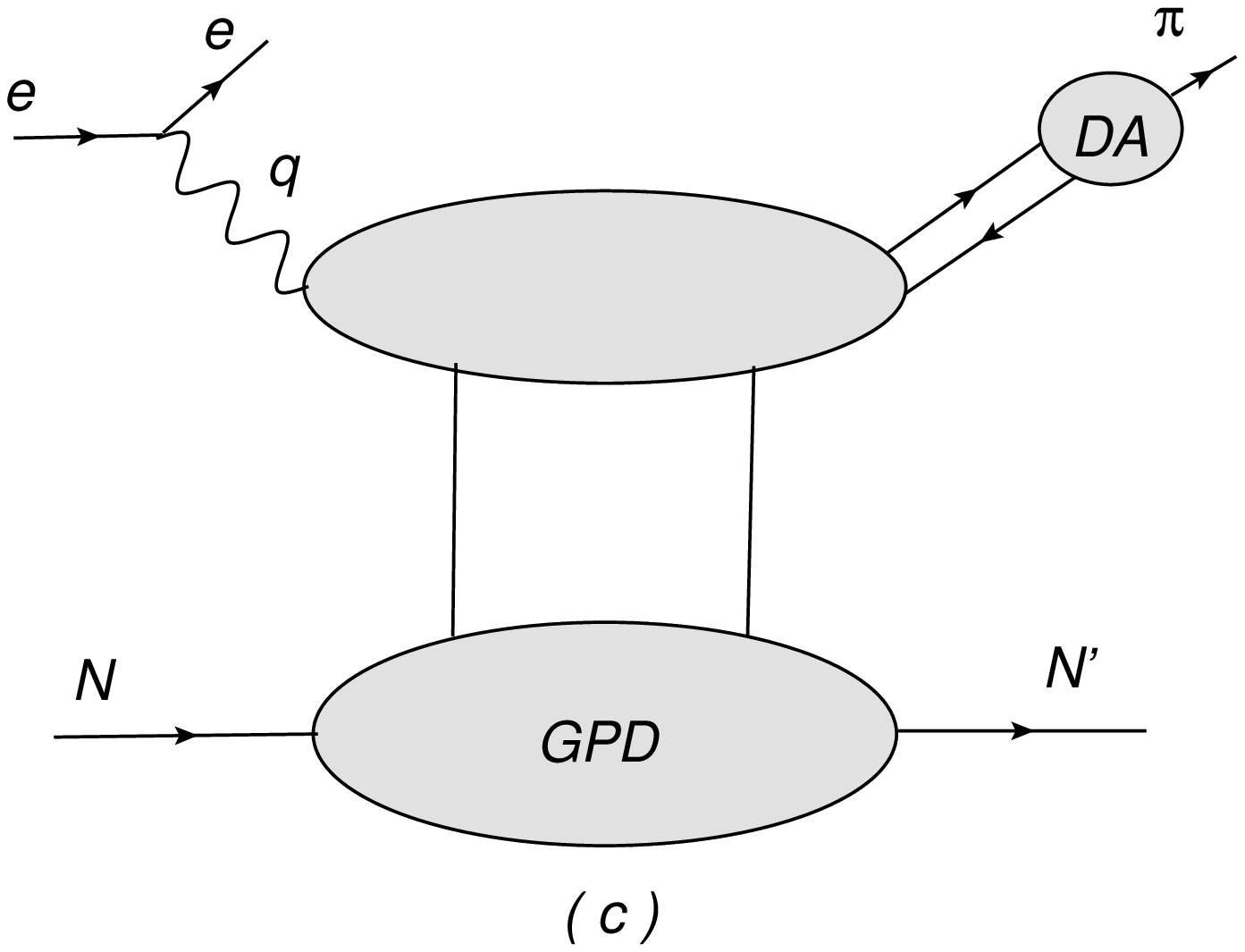} ~~~\includegraphics[width=3.8cm]{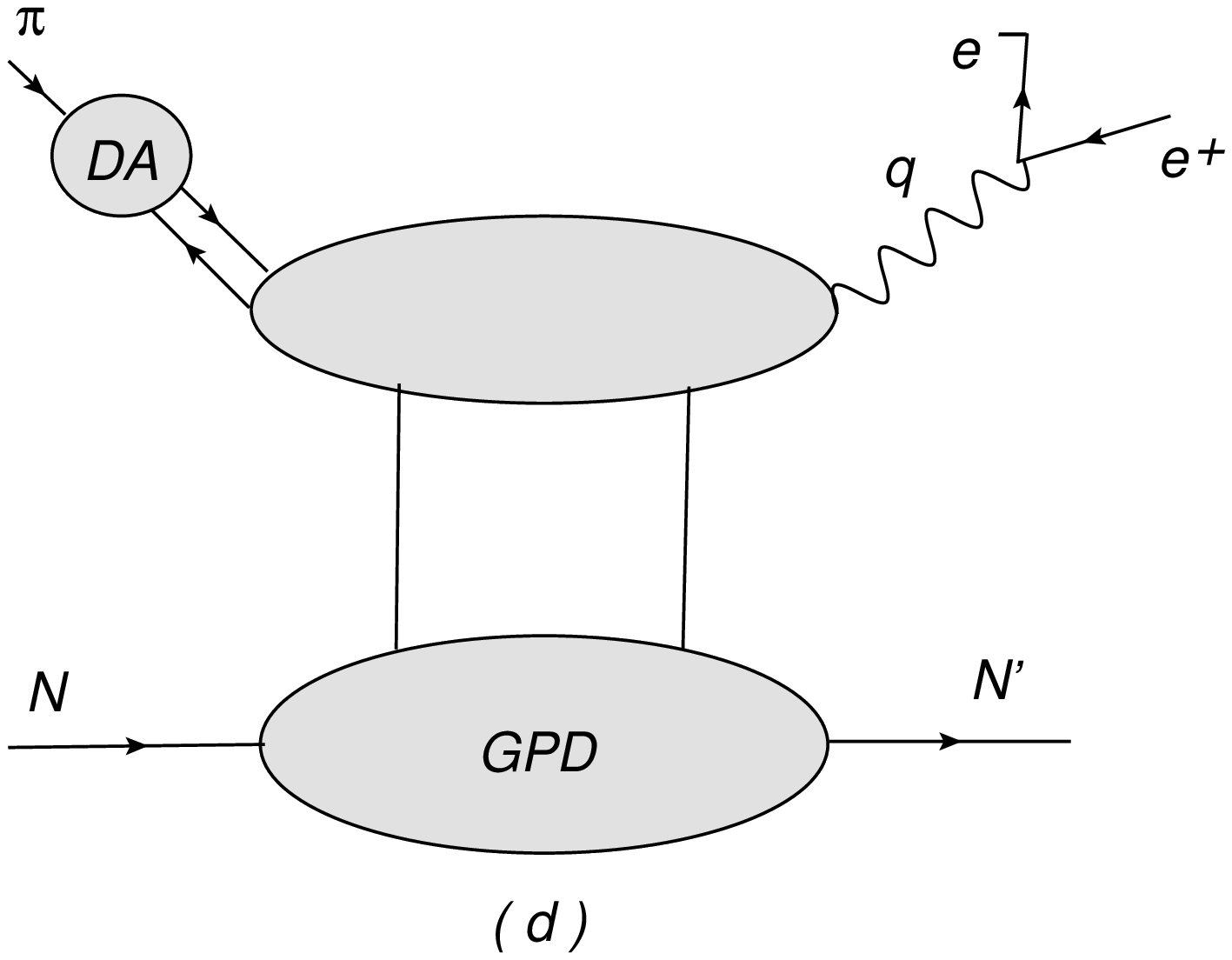}
\vspace{-2mm}
\caption{The DVCS (a) and TCS (b) processes, as well as meson electroproduction (c) and exclusive Drell Yan in $\pi N$ collisions (d) are linked by time reversal and analyticity. They factorize in hard coefficients (upper blob), generalized parton distributions (lower blob) and
%possibly
distribution amplitudes (c,d).}
\label{fig1}
\end{figure}

We shall detail two instances of direct interest to near future phenomenological studies, illustrated in Fig.\ref{fig1}, firstly
near forward deeply virtual Compton scattering (DVCS) and timelike Compton scattering  (TCS), and secondly near forward deeply virtual meson leptoproduction  (DVMP) and mesoproduction of a lepton pair.
The momentum transfer square $t$ in these processes is taken to be small w.r.t.~the large virtuality of one  photon.

%%%%%%%%%%%%%%%%%%%%%%%%%%%%%%%%%%%%%%%%%%%%%%%%%%%%%%%%%%%%%%%%%%%%%%%%%
\unitlength 1mm
%%%%%%%%%%%%%%%%%%%%%%%%%%%%%%%%%%%%%%%%%%%%%%%%%%%%%%%%%%%%%%%%%%%%%%%%%

\vspace{.2cm}
\paragraph*{The DVCS and TCS amplitudes.}
Let us begin with  near forward virtual Compton scattering
\begin{equation}
\gamma^{(*)}(q_{in}) N(p) \to \gamma^{(*)}(q_{out}) N'(p')\,.
\end{equation}
In its spacelike version the DVCS amplitude is accessible in deep electroproduction of a  photon, i.e., $q_{out}^2 =0$,
\begin{equation}
e(k_1) N(p) \to e'(k_2) \gamma(q_{out}) N'(p')
\end{equation}
with a large spacelike virtuality $q_{in}^2 = (k_1-k_2)^2=-{\cal Q}^2$ \cite{historyofDVCS}.
The  timelike TCS amplitude is accessible in the photoproduction, i.e., $q_{in}^2 =0$, of a lepton pair \cite{BDP}:
\begin{equation}
\gamma(q_{in}) N(p) \to l^+(k^+) l^-(k^-) N'(p')
\end{equation}
with a large timelike virtuality $q_{out}^2 =(k^+ + k^-)^2=+ {\cal Q}^2$.
The other common variables, describing the processes of interest in this generalized Bjorken limit, are the scaling variable $\xi$ and skewness $\eta  > 0$:
\begin{equation}
\xi = -\frac{q^2_{out}+q^2_{in}}{q^2_{out}-q^2_{in}} \eta\,, \quad
\eta =\frac{q^2_{out}-q^2_{in}}{(p+p')\cdot(q_{in}+q_{out})}\,.
\label{eq:skewnessdef}
\end{equation}
Hence, $\xi = +\eta >0$ in DVCS and  $\xi = -\eta < 0$  in TCS kinematics. This allows us to relate spacelike and timelike amplitudes for equal $\eta$, $t$, and ${\cal Q}^2$ values by the rule:
\begin{eqnarray}
{\cal F}(\xi=\eta,t,{\cal Q}^2) \stackrel{\mbox{\tiny SL} \to \mbox{\tiny TL}}{\Longrightarrow} {\cal F}(\xi=-\eta,t,-{\cal Q}^2)\,,
\label{rules-SL2TL}
\end{eqnarray}
where the  c.o.m.~energy square $s = (p + q_{in})^2$ might differ.

We first study the  DVCS amplitude which is usually parameterized in terms of Compton form factors (CFFs) \cite{gpdrev}. After  renormalization,
a leading twist CFF read as sum over quarks ($q$) and gluon ($g$)  in its factorized form:
\begin{eqnarray}
%\mathcal{A}^{\mu\nu} = -g_T^{\mu\nu}
{\cal F}(\xi, t,{\cal Q}^2) =
\int_{-1}^1\!\! dx\!
%\left[\!
\sum_{i=u,d,\cdots, g}\!\! {^ST^i(x,\xi)} {F^i(x,\xi,t,\mu^2)} %+ {^TT^g(x)} F^g(x)\!
%\right]
,
\label{eq:factorizedamplitude}
\end{eqnarray}
where  we adopt for generalized parton distributions (GPDs) $F^i$ the common conventions \cite{gpdrev}.
The hard coefficients $^ST^i$ depend on both the virtuality and factorization scale and read to next-to-leading order (NLO) accuracy in $\alpha_s$:
\begin{eqnarray}
 {^ST^i}\stackrel{\rm NLO}{=} {^SC_{0}^i} + \frac{\alpha_s(\mu^2)}{2\pi}\left[
 {^SC_1^i} + {^SC_{coll}^i} \ln\frac{{\cal Q}^2}{\mu^2}\right].
\label{eq:coefficients}
\end{eqnarray}
Note that the possible distinction between factorization scale $\mu_F$ and renormalization scale $\mu_R$ has no consequence on our arguments and we simplify notations by equating $\mu=\mu_F=\mu_R$. The collinear coefficient $^SC_{coll}^i$ are given as convolution of the Born term $^S{C_0^i}$ with the GPD evolution leading order (LO) kernels. Since DVCS amplitude is symmetric under $s\leftrightarrow u$-channel crossing, the CFFs and $^SC_{\cdots}^i$ coefficients have definite symmetry properties under $\xi$-reflection.  Moreover, boost-invariance tells us that all $^SC_{\cdots}^i$ coefficients are functions of the variable $x/\xi$, apart from an overall scaling factor. For $\xi$ (or $s\leftrightarrow u$) symmetric coefficients we write here explicitly
\begin{equation}
C_0^q(x,\xi)= e_q^2\left( \frac{1}{\xi-x} -  \frac{1}{x+\xi} \right),\quad  C_0^g(x,\xi)= 0\,.
\label{qC0}
\end{equation}
Note that gluons  do not contribute in LO but at NLO:
\begin{eqnarray}
C_{coll}^g(x,\xi)&=&  - \frac{\frac{1}{2}\sum_q e_q^2}{(\xi+x)^2} \ln\frac{\xi-x}{2\xi} + (x \rightarrow -x )\,,
\label{gCcoll}
\\
 C_1^g(x,\xi) &=& \frac{\frac{1}{2}\sum_q e_q^2}{(\xi+x)^2}
\left(\frac{3\xi-x}{\xi-x}  -  \frac{1}{2}\ln\frac{\xi-x}{2\xi}  \right)
\label{gC1} \\
&& \times 
\ln\frac{\xi-x}{2\xi} +(x \rightarrow -x)
\,, \nonumber
\end{eqnarray}
where the remaining quark and antisymmetric coefficients in this representation can be read off from \cite{BelMueNieSch00}. Obviously,
for $z=x/\xi$ these functions are holomorphic in the complex plane except for a $s$-(and $u$)-channel poles at $z=1$ ($z=-1$)  and $s$-(and $u$-)channel cuts $[1,\infty]$ ($[-\infty,-1]$) on the real axis. Their physical value on the cuts (or poles) is governed by
causality, i.e., by the $+i\epsilon$ prescription of propagators. This yields the extension of the scaling variable $\xi_S =\xi -i\epsilon$  into the complex domain, which can be also read off from $\xi={\cal Q}^2/(2s+{\cal Q}^2)$, resulting from (\ref{eq:skewnessdef}), and decorating $s$ with
$+i\epsilon$.  All hard coefficients in (\ref{eq:coefficients}) can be then uniquely extended:
\begin{eqnarray}
^SC_{\cdots}^i(x,\xi) = C_{\cdots}^i(x,\xi_S)\,,
\label{qdvcscoefficients}
\label{gdvcscoefficients}
\end{eqnarray}
consistent with the physical sheet.

For the TCS amplitude, i.e., time-like CFFs, the situation is in general more intricate due to existence of possible poles and cuts, caused by the time-like virtuality of the outgoing photon. However, in our perturbative description of TCS we  require that we are away from the resonance region and we might employ the substitution rule (\ref{rules-SL2TL}) and causality to find the hard coefficients in the timelike region by analytic continuation, e.g.,  from (\ref{qC0}--\ref{gC1}). However, from (\ref{eq:coefficients}) we immediately see that the factorization
$\ln\frac{{\cal Q}^2}{\mu^2}$ goes into $\ln\frac{-{\cal Q}^2}{\mu^2}$, providing us additional $\pm i\pi C^i_{coll}$  terms at NLO. To pick up the proper sign, we might analyze Feynman diagrams or, equivalently, we can use a convolution representation for DVCS coefficients, in which the $\ln\frac{{\cal Q}^2 (\xi -i\epsilon \mp x)}{2\xi \mu^2}$  appears \cite{Belitsky:1997rh}. As we show below, the rule (\ref{rules-SL2TL}) together with the $i\epsilon$ prescription provides then an unique answer.

Let us verify this statement and also provide us a more usable timelike-to-spacelike relation.
At Born level we easily realize  in accordance with a diagrammatic evaluation a rule, conveniently written with $\xi_T=\eta+i\epsilon$:
\begin{eqnarray}
\hspace{-0.5cm}
\frac{1}{\xi -i \epsilon\mp x} \stackrel{\mbox{\tiny SL} \to \mbox{\tiny TL}}{\Longrightarrow}
 \frac{1}{-\eta -i\epsilon \mp x} =\frac{-1}{\xi_T\pm x} =
\frac{-1}{\xi_S^\ast \pm x}\,.
\label{rule:DVCS2TCS}
\end{eqnarray}
This exercise exemplifies our main  result, namely,  the {\em timelike} $s (u)$-channel coefficients are given by {\em complex conjugation} of the {\em spacelike} $u (s)$ one. Utilizing Schwarz reflection principle, we  write for a generic (N)LO coefficient:
\begin{eqnarray}
{^SC}(x,\xi_S) \stackrel{\mbox{\tiny SL} \to \mbox{\tiny TL}}{\Longrightarrow}
{^TC}(x,\xi_T) = \mp\, {^SC^{\ast}}(-x,\xi_S)\,,
\label{ruleS2Ts2u}
\end{eqnarray}
where the upper sign applies for quarks and the lower for gluons (compared to quark GPDs our gluon GPDs contain a relative $x$ and so
quark and gluon coefficients have different symmetry properties under $\xi$- and $x$-reflection).
From the analyticity of hard coefficients, see, e.g., (\ref{gCcoll},\ref{gC1}), and the substitution (\ref{rules-SL2TL}) we also establish the rule (\ref{ruleS2Ts2u}) at NLO. As said, there is an additional imaginary part, uniquely fixed by causality, that is associated with the factorization logarithms ($\ln$'s). Indeed, in a diagrammatic NLO calculation \cite{Pire:2011st}  we realize that they appear in $\ln\frac{-\hat{s}-i\epsilon}{\mu^2}$ and $\ln\frac{-\hat{u}-i\epsilon}{\mu^2}$ terms, where $\hat{s}=\frac{x-\xi}{2\xi}{\cal Q}^2$ and $\hat{u}=-\frac{\xi+x}{2\xi}{\cal Q}^2$ are Mandelstam variables for partonic subprocesses. In the DVCS case the $\hat s$ cut is contained in:
\begin{eqnarray}
\ln\frac{-\hat{s}_S-i\epsilon}{\mu^2} = \ln\frac{{\cal Q}^2}{2\xi\, \mu^2} + \ln(\xi-i\epsilon-x)\,,
\end{eqnarray}
which after applying (\ref{rules-SL2TL}) goes into the TCS  expression, which can then be expressed by the spacelike $u$-channel
contribution and a $-i\pi$ addendum:
\begin{eqnarray}
\ln\frac{-\hat{s}_T-i\epsilon}{\mu^2} &=& \ln\frac{{\cal Q}^2}{2\eta\, \mu^2} + \ln(-\eta-i\epsilon-x)
\\
&=&
\left[\ln\frac{-\hat{u}_S-i\epsilon}{\mu^2} \right]^\ast - i \pi\,.
\nonumber
\end{eqnarray}
An analogous result holds for the $\hat{u}$-channel and, thus, independently from the considered channel the
space--to--timelike relation (\ref{ruleS2Ts2u}) is accompanied by
\begin{eqnarray}
\ln\frac{{\cal Q}^2}{\mu^2} \stackrel{\mbox{\tiny SL} \to \mbox{\tiny TL}}{\Longrightarrow}
\ln\frac{{\cal Q}^2}{\mu^2} -i \pi\,.
\label{ruleS2Tln}
\end{eqnarray}

Employing the space--to--timelike relation (\ref{ruleS2Ts2u},\ref{ruleS2Tln}) to the net NLO coefficient (\ref{eq:coefficients}) we find the timelike ones:
\begin{equation}
{^T}T^i \stackrel{\rm NLO}{=} \pm {^S}T^{i\,*} \mp i\pi \frac{\alpha_s}{2\pi} {^S}C_{coll}^{i\,*}\,; %+ {\cal O}(\alpha_s^2)\
\label{fullrelation}
\end{equation}
%where
upper (lower) sign applies to $\xi$-(anti)sym\-metric CFFs.

For the symmetric case the space--to--timelike relation (\ref{fullrelation})
has been exemplified by a diagrammatic NLO evaluation  \cite{Pire:2011st}.
As we have seen,  (\ref{fullrelation}) arises from general field theoretical principles
and is an example of a more general result for hard NLO coefficients at
twist-two accuracy.

%%%%%%%%%%%%%%%%%%%%%%%%%%%%%%%%%%%%%%%%%%%%%%%%%%%%%%%%%%%%%%%%%%%%%%%%%

\vspace{.2cm}
\paragraph*{DVMP and exclusive Drell-Yan.}
Let us now turn to a slightly different pair of reactions where amplitudes factorize in both GPDs and a meson distribution amplitude (DA). Specifically, we consider $\gamma_L^* N \to \pi N^\prime$, a subprocess in near forward  leptoproduction,  and
$\pi N \to \gamma_L^*  N^\prime$, appearing in the exclusive limit of Drell--Yan process. The factorization theorem \cite{Collins:1996fb} states that the  $ \gamma_L^* p \to \pi^+ n$ amplitude, written in terms of $\widetilde{\cal F}_{\pi^+}(\xi,t,{\cal Q}^2)$ transition form factors (TFFs),
factorizes up to a constant factor as
\begin{eqnarray}
\widetilde{\cal F} \propto\frac{1}{\cal Q}\!
\int_{0}^{1}\!\!\!du\!\! \int_{-1}^{1}\!\!\!  dx\,\widetilde{F}^{ud}(x,\xi,t){^S T_{ud}}(u,x,\xi)\varphi_\pi(u)\,.
\label{tffpi+}
\end{eqnarray}
Here, $ud$ denote the exchanged quark pair, the flavor off-diagonal GPD  $\widetilde F^{ud} = \widetilde F^u-\widetilde F^d$ is expressed by diagonal ones via $SU(2)$ symmetry, and the pion DA $\varphi_\pi$ is symmetric w.r.t.~$u \to 1-u$.
In analogy to DVCS,  we  introduce  $C(u,x,\xi)$ coefficients and write
\begin{eqnarray}
{^ST}_{ud} =  \left[e_u C(u,x,\xi_S)  - e_d C(u,-x,\xi_S)\right],
\label{Tud}
\end{eqnarray}
where the physical sheet is picked up by $-i\epsilon$ in $\xi_S$.
Note that we use here and in the following $u \to 1-u$ symmetry and that
already the LO result is proportional to $\alpha_s(\mu^2)$,
\begin{eqnarray}
\hspace{-0.6cm}
&& C\stackrel{\rm NLO}{=} \alpha_s(\mu^2) C_0 + \frac{\alpha^2_s(\mu^2)}{2\pi}\left[ C_{div} \ln\frac{{\cal Q}^2}{\mu^2}  + C_{1}\right],
\label{DVMP-C}
\\
&& C_0(u,x,\xi)= \frac{1}{u (\xi-x)}\,,
\label{DVMP-C0}
\\
&& C_{div}=\frac{-\beta_0}{2}  C_0 +  C^{F}_{coll} +  C^{\varphi}_{coll}\,.
\label{DVMP-Cdiv}
\end{eqnarray}
Here, $\beta_0 = 11-2n_f/3$ controls the running of $\alpha_s$ at LO, the collinear coefficients $C^{F}_{coll}$ and
  $C^{\varphi}_{coll}$ are given as convolution of LO evolution kernels with the LO coefficient (\ref{DVMP-C0}). 
 All these coefficients can be obtained from  known pion form factor results \cite{Belitsky:2001nq}. 
 $C_{\cdots} $ are  $ {\cal Q}^2$ independent. 
  Moreover, analytic properties, seen in DVCS coefficients such as  (\ref{qdvcscoefficients}), hold  for the coefficients in (\ref{DVMP-C}) as function of $z=x/\xi$, too, which justifies the replacement $\xi\to \xi_S$ in (\ref{Tud}) \cite{Pasprivate}.

The factorization proof \cite{Collins:1996fb} may be extended  to the crossed reaction $\pi N \to \gamma_L^* N'$. The time-like TFFs ${^T\widetilde{\cal F}_{\pi^-}}$, appearing in  $\pi^- p \to \gamma_L^* n$, might be in full analogy to the space like form factor (\ref{tffpi+}) written as convolution of $\widetilde{F}^{du}= -\widetilde{F}^{ud}$ GPD and pion DA, where
hard coefficients read to LO accuracy as \cite{BDPpi}:
\begin{eqnarray}
\hspace{-5mm}
{^TT_{du}}(u,x,\xi_T) = \left[e_u C_0(u,x,\xi_T)  - e_d C_0(u,-x,\xi_T)\right].
\label{^TTdu}
\end{eqnarray}

Taking the physical sheet in spacelike region, the reflection (\ref{rules-SL2TL}) implies the space--to--timelike relation
(\ref{ruleS2Ts2u}) for NLO coefficients. As in DVCS, from the explicit NLO result we can read of the rule  (\ref{ruleS2Tln})
for the continuation of renormalization and factorization $\ln$'s, e.g., the $\ln$'s of the $\beta_0$ proportional part can
be collected in  a $\ln\frac{-u \hat{s}-i\epsilon}{\mu^2}$ or $\ln\frac{-u \hat{u}-i\epsilon}{\mu^2}$ term.
Hence, both rules can be employed to the net coefficient (\ref{Tud}) and so
 we obtain with  (\ref{DVMP-C}--\ref{DVMP-Cdiv}) the NLO approximation for timelike coefficient (\ref{^TTdu}), where
\begin{eqnarray}
\hspace{-3mm}
{^T C}(u,x,\xi_T) \stackrel{\rm NLO}{=} - \left[C^\ast -i\pi  \frac{\alpha_s}{2\pi}C^\ast_{div}\right](u,-x,\xi_S)\,.
\label{relationC-SL2TL_2}
\end{eqnarray}
Note that  coefficients for spacelike [timelike] TFFs $\widetilde{\cal F}_{\pi^-}$  [${^T\widetilde{\cal F}}_{\pi^+}$] for $\pi^-$ DVMP [and exclusive Drell--Yan in $\pi^+$] off neutron follows from (\ref{Tud}) [(\ref{^TTdu})] by $u\leftrightarrow d$ exchange.

This result generalizes the relation obtained in Ref. \cite{BRS} between the  timelike and spacelike pion form factors,  in which only the first and third term on the r.h.s. of Eq. \ref{DVMP-Cdiv} appear.

\vspace{.2cm}
\paragraph*{Phenomenological perspectives.}
As we have seen, the space--to--timelike relation of hard coefficients is at NLO modified by  $-i\pi$ proportional terms that are  associated with factorization and renormalization $\ln$'s. Since GPDs and DA are real valued, our findings imply a relation among CFFs or TFFs. In the
case of $\xi$-(anti)symmetric CFFs, called  $\cal H$ ($\widetilde{\cal H}$) and $\cal E$ ($\widetilde{\cal E}$), Eq. (\ref{fullrelation}) yields, e.g.,
\begin{eqnarray}
{^T{\cal H}}  &\stackrel{\rm NLO}{=}&  {\cal H}^\ast - i\pi\, {\cal Q}^2\frac{\partial}{\partial {\cal Q}^2} {\cal H}^\ast\,,
\label{cffH2cffTH}
\\
{^T\widetilde{\cal H}}  &\stackrel{\rm NLO}{=}&  -\widetilde{\cal H}^\ast + i\pi\, {\cal Q}^2\frac{\partial}{\partial {\cal Q}^2} \widetilde{\cal H}^\ast\,.
\label{cffHt2cffTHt}
\end{eqnarray}
An analog relation  connects (up to a conventional phase) timelike $\pi^\pm$  with spacelike $\pi^\mp$ TFFs, see (\ref{tffpi+},\ref{Tud},\ref{^TTdu},\ref{relationC-SL2TL_2}): 
\begin{eqnarray}
{^{\rm T}\widetilde{\cal H}_{\pi^\pm}} \stackrel{\rm NLO}{\simeq}  {\widetilde{\cal H}_{\pi^\mp\,}^\ast} - i\pi {\cal Q} \frac{ \partial }{\partial {\cal Q}^2}\; {\cal Q}\,{\widetilde{\cal H}_{\pi^\mp\,}^\ast} 
\,.
\label{tffHt2tffTHt}
\end{eqnarray}

The NLO relations (\ref{cffH2cffTH}--\ref{tffHt2tffTHt}) tell us that if scaling violations are small, the timelike CFFs (TFFs) can be obtained from the spacelike ones by complex conjugations. Moreover, GPD model studies indicate that in the valence region, i.e., for $\xi \sim 0.2$,   CFFs  might only evolve mild. This rather generic statement, which will be quantified by model studies \cite{Moutarde}, might be  tested in future (after 12GeV upgrade) Jefferson Lab experiments.

\begin{figure}
\includegraphics[width=7cm]{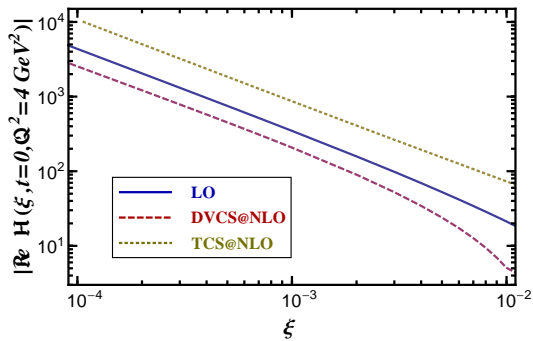}
\vspace{-3mm}
\caption{The real part of CFF $\mathcal{H}$ vs.~$\xi$ with $\mu^2=Q^2= 4 \textrm{~GeV}^2$  and $t=0$ at LO (solid) and NLO  for DVCS (dashed).
For  TCS at NLO  its negative value is shown as dotted curve. }
\label{fig:Re}
\end{figure}
On the other hand it is known that the evolution of  CFF $\cal H$ in the small $\xi$ region is driven by the ``pomeron'' pole in the gluon evolution
kernel which also interfere with the effective ``pomeron'' intercepts of  GPDs  at the input scale.
%For the sake of illustration, let us just present a specific example of the effect of the NLO corrections in DVCS and TCS processes. All observables are %expressed through the Compton Form Factors (CFFs) defined as the convolution of coefficient functions with appropriate generalized parton distributions %\cite{BDP}. Let us concentrate on the CFF $\mathcal{H}$. where $H^q$ and $H^g$ are the appropriate GPDs.
The effective ``pomeron'' trajectory induces then that the imaginary part $\Im{\rm m}{\cal H}$ dominates over the real one $\Re{\rm e}{\cal H}$, which
is consistent with a phenomenological analysis of HERA data \cite{Kumericki:2009uq}.
Since of the $-i\pi$ proportional NLO addenda in (\ref{cffH2cffTH}), the small $\Re{\rm e}{\cal H}$
will only mildly influence the LO prediction $\Im{\rm m}{^T{\cal H}} \stackrel{\rm LO}{=} -\Im{\rm m}{\cal H}$. On the other hand we expect huge  NLO corrections
to $\Re{\rm e}{^T{\cal H}} \stackrel{\rm LO}{=} \Re{\rm e}{\cal H}$, induced by $\Im{\rm m}{\cal H}$.
%DVCS and TCS have rather similar effects on the imaginary parts, diminishing its absolute value. The situation is very different for the real part where we %observe huge differences between NLO DVCS and NLO TCS corrections.
Utilizing Goloskokov-Kroll model
for $H$ GPDs \cite{Goloskokov:2006hr}, we illustrate this effect
in Fig.~\ref{fig:Re} for $10^{-4}\le  \xi \le 10^{-2}$, accessible in a suggested Electron-Ion-Collider \cite{Boer:2011fh}, and $t=0$.  We plot $\Re{\rm e}{\cal H}$ vs.~$\xi$, for LO DVCS or TCS (solid), NLO DVCS (dashed) and NLO TCS (dotted) at
the input scale $\mu^2={\cal Q}^2 = 4 \textrm{~GeV}^2$. In the case of NLO TCS $-\Re{\rm e}{^T{\cal H}}$ is shown, since even the sign changes. We read off that the NLO correction to $\Re{\rm e}{^T{\cal H}}$ is of the order of $-400\%$ and so
%That fact is directly connected to the  $i\pi$ term present in equation (\ref{fullrelation}), which indeed significantly modifies the  phase of the amplitude.
the real part in TCS becomes of similar importance as the imaginary part. This NLO prediction is testable via a
lepton-pair angle asymmetry, governed by  $\Re{\rm e}{^T{\cal H}}$ \cite{BDP}. Such drastic effect of the timelike nature of outgoing photon was also found in the dipole model approach \cite{Schafer:2010ud}.

\vspace{.2cm}
\paragraph*{Conclusions.}
\noindent
We have shown that the factorization property of exclusive amplitudes at leading twist together with analyticity  allow to link various processes at
NLO accuracy.  Thereby, we specialized to near forward  processes in the generalized Bjorken regime where collinear factorization holds. The space--to--timelike relation (\ref{ruleS2Ts2u},\ref{ruleS2Tln}) helps to understand the previously published result of \cite{Pire:2011st}, leads to new results written in (\ref{fullrelation},\ref{relationC-SL2TL_2}), and indicates a more general relation that might be established by a perturbative analyze of Feynman diagrams.

The extension of $q \bar q$ and $g g$ exchange to $q q q$ exchange \cite{Frankfurt:1999fp} in a generalized Bjorken regime, much related to the  DVCS one, generalizes the GPD concept, yielding the definition of transition distribution amplitudes (TDAs) and to a factorized formula for backward DVCS  and backward leptoproduction of a $\pi$ meson  \cite{Pire:2004ie}. For the latter  $\pi N$ TDAs factorize from the hard subprocess. The corresponding timelike processes occur in meson proton scattering into a massive lepton pair and nucleon. Here also analyticity allows to relate NLO corrections in both processes. We shall discuss that elsewhere.

%We also did not discuss  higher twist contributions to these processes since their factorization properties are not established. Nor we  considered the simpler case of electromagnetic form factor, discussed earlier in \cite{BRS,GP}, neither fixed angle scattering,   which amplitudes are known to be  infrared sensitive \cite{Botts:1989kfRalston:1982pa} and necessitate the inclusion of Sudakov like resummation \cite{Sud}.

\vspace{.2cm}
\paragraph*{Acknowledgements}
\noindent
We are grateful to Tolga Altinoluk, Igor Anikin, Markus Diehl, Thierry Gousset, Vadim Guzey, Herv\'e Moutarde, Kornelija Passek-Kumeri{\v c}ki,  Anatoly Radyushkin, John P. Ralston,  Franck Sabati\'e, Oleg Teryaev,  Samuel Wallon and Christian Weiss  for useful discussions and correspondence.
This work is partly supported by the Polish Grant NCN No DEC-2011/01/D/ST2/02069, the French-Polish collaboration agreement Polonium,
BMBF grant no. 06BO9012, and the Joint
Research Activity "Study of Strongly Interacting Matter"
(acronym HadronPhysics3, Grant Agreement n.283286) under the
Seventh Framework Programme of the European Community.

%%%%%%%%%%%%%%%%%%%%%%%%%%%%%%%%%%%%%%%%%%%%%%%%%%%%%%%%%%%%%%%%%%%%%%

\end{document}